\pdfoutput=1
\documentclass[conference]{IEEEtran}
\IEEEoverridecommandlockouts

\usepackage{cite}
\usepackage{amsmath,amssymb}
\usepackage{graphicx}
\usepackage{booktabs}
\usepackage{array}
\usepackage[htt]{hyphenat}
\usepackage{xcolor}
\usepackage{url}
\usepackage[hidelinks]{hyperref}
\usepackage{float}
\begin{document}

\title{Beyond Maliciousness: Static and Runtime Attack-Surface Profiling of Agent Skills}

\author{
\IEEEauthorblockN{
Ratnesh Pandey\textsuperscript{a},
Japneet Singh\textsuperscript{a},
Harsh Akshit\textsuperscript{a},
Sonali Tyagi\textsuperscript{b},
Cassie Crossley\textsuperscript{c}
}
\IEEEauthorblockA{\textsuperscript{a}Metano Labs\\}
\IEEEauthorblockA{\textsuperscript{b}Purdue University\\}
\IEEEauthorblockA{\textsuperscript{c}VulNow\\}
}
\maketitle

\begin{abstract}
Agent skills are Markdown documents that a tool-using agent reads as
instructions and then acts on with the privileges of the session that loaded them. Installing one grants authority (the standing to direct the agent, not new privilege), and the size of that grant is what we
measure.

We present a two-phase attack-surface profiling methodology that combines whole-bundle static analysis and dependency provenance checking with instrumented sandbox execution and separate model-based assessment of the resulting traces. We apply the methodology to 6{,}309 listings from two marketplaces, selecting listings for execution without conditioning that decision on a prior threat or risk signal. Half the corpus carries evidence of shell-execution capability; the corresponding shares are 35\% for filesystem read, 31\% each for filesystem write and network egress, and 13\% for credential access; seven listings in ten carry at least one elevated capability. That is the surface a user accepts at install time, and almost none of it is visible then.

A permission model exists but is close to inert. Authors may list the tools a skill pre-approves in its own frontmatter, which removes confirmation prompts rather than restricting what the skill can reach; of the 6{,}061 listings whose
frontmatter parses, 770 declare anything at all, the vocabulary cannot express
credential or environment access, and nothing checks what it does express against the bundle.
Declaring is also not declaring accurately: 59\% of declaring listings
understate what their bundle can do. ClawHub declares roughly twice as often as skillsmp, yet under-declares on 88\% of its declarations against 31\%.

Triage at the marketplace boundary therefore cannot rest on what a listing declares. It must rest on what the bundle can do. These results position capability-oriented attack-surface profiling as a necessary complement to maliciousness detection at the marketplace boundary.
\end{abstract}

\section{Introduction}
\label{sec:why}

An agent skill is a directory holding a Markdown document, \texttt{SKILL.md},
together with whatever scripts and data it chooses to ship alongside. When an agent loads a skill, the document's
contents become instructions the agent follows, and the scripts become programs
the agent runs. Installation is a copy: a user places the directory where the
agent looks for it, and from that point the skill executes inside the agent's
existing session, with the shell, the filesystem, the network reach and the
credentials the agent already holds. A permission-shaped field does exist: an author may list tools in the skill's frontmatter. The specification defines them as pre-approved to run~\cite{agentskills2026spec}, and in at least one major agent the list removes confirmation prompts for those tools while leaving every other tool callable~\cite{claudecode2026skills}, and a pre-approval so declared travels into every repository that installs the skill~\cite{scanningharness2026}. It is a grant, not a restriction. It is opt-in; it is written by the same author who wrote the skill, nothing checks it against what the bundle actually contains, and the only field that touches credentials or the environment is an optional free-text compatibility string that no analyzer can check a bundle against. 

A skill operates within several layers of control. The operating system bounds the privileges of the agent process, while the agent harness determines which tools and resources are available and how their use is authorized ~\cite{agentskills2026spec, skillguard2026permissioncentric}. The skill contributes instructions and may include scripts that the agent invokes~\cite{agentskills2026spec}. The Agent Skills specification defines allowed-tools as an optional list of pre-approved tools, not as a required statement of every capability the bundle may exercise~\cite{agentskills2026spec}. Even an accurate declaration therefore cannot establish the limits of actions performed through a shell or an invoked script. Assessing a skill requires examining what its bundle can bring into play under the controls of the session that loads it, as well as what its author declares.

Capability profiling addresses one part of this broader security
problem: identifying what a bundle may bring into play before
deployment. This information can inform marketplace review and
the selection of harness restrictions, but it does not establish
that the agent will use those capabilities safely. Safe operation
also depends on authorization policies, isolation, and controls
on actions during execution.

\begin{figure}[t]
\centering
\includegraphics[width=\columnwidth]{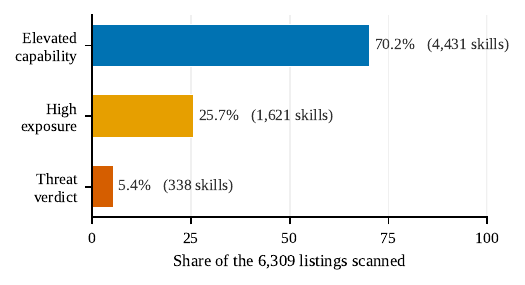}
\caption{Three properties of the same 6{,}309 marketplace listings. Most listings carry at least one consequential capability, a quarter are judged to carry serious exposure, and one in twenty carries a threat verdict. Capability is inferred from the bundle; intent is a thresholded verdict.}
\label{fig:exposure}
\end{figure}

Adoption has been fast. A December 2025 crawl of one marketplace found
15{,}082 skills~\cite{liu2026agentskillswildempirical}; a crawl of the same
marketplace one month later found 73{,}193~\cite{liu2026donotmention}; and by
March 2026 a study spanning three distribution platforms and GitHub counted
238{,}180 unique skills~\cite{holzbauer2026contextmatters}. Skills are written
quickly, copied between repositories and revised in
place~\cite{gao2026registrytorepository}, and installation asks the user to
approve nothing beyond the copy itself. What arrives is trusted by default, and
the evidence that the trust is sometimes misplaced is now substantial: a static
study of 31{,}132 skills reported that roughly one in four carried a
vulnerability pattern and one in twenty showed likely malicious
intent~\cite{liu2026agentskillswildempirical}, a detonation study confirmed 157
malicious skills in the wild~\cite{liu2026donotmention}, and practitioner
reporting has documented live malicious campaigns on a public
registry~\cite{snyk2026toxicskills}.

Those results answer one question: Is this skill trying to cause harm? They do not answer a second and, for most users, more consequential question: What authority does installing the skill grant? A skill that is entirely benign in intent can still expand the set of actions available to an agent operating with the user's privileges. In the corpus studied here, half the listings carry evidence of shell-execution capability, and seven in ten carry at least one elevated capability. Filesystem reads appear in 35\% of listings, filesystem writes and network access in 31\% each, and credential access in 13\%. Fig.~\ref{fig:exposure} sets this capability footprint against the much smaller population carrying a threat verdict. Most of this authority is not declared in a form that a user or marketplace can inspect before installation. The exposure associated with installing a skill is therefore consequential but largely invisible at the point of decision.

\textbf{Terminology.} We report three distinct listing-level quantities. \textit{Threat} concerns evidence of malicious intent. \textit{Attack surface} describes the capability footprint inferred from a listing, where an individual \textit{capability} is an action that the bundle of evidence indicates may be available, such as shell execution, filesystem access, network egress, or credential access. \textit{Risk} describes the severity of the resulting exposure, based on the reach of that capability footprint and associated amplification factors. Capability, attack surface, and exposure therefore do not, by themselves, imply malicious intent, observed exploitation, or realized harm. \textit{Authority.} By `grants authority' we mean that installation gives a skill the standing to direct the agent, persistently and without per-action confirmation ~\cite{li2026towards}. It grants no privilege the session did not already hold. What a skill can reach is the agent's existing ambient authority ~\cite{miller2003capability}, and a skill that turns it against the user makes the agent a \emph{confused deputy} ~\cite{hardy1988confused}. The quantity we measure is how much of that existing authority a skill brings into play. Section~III defines how each quantity is operationalized.

\section{Related Work and Contributions}
\label{sec:why-new}

\subsection{Threat Measurement and Its Limits}

The security of agent skills is now well studied, and the studies agree on what
to report: a verdict on whether a skill is trying to cause harm. Static analysis
with model assistance has been applied to 31{,}132
skills~\cite{liu2026agentskillswildempirical} and, with repository context, to
238{,}180~\cite{holzbauer2026contextmatters}. Detonation has confirmed 157
malicious skills in the wild~\cite{liu2026donotmention}. Vendor research has
documented live campaigns on a public registry~\cite{snyk2026toxicskills}.

One study comes close to the distinction this paper is built on. Comparing three
scanners over 67{,}453 skill versions, Koc et al. observe that a skill may
be benign in content and still risky if it grants excessive authority, and that
a skill carrying high agentic risk may be entirely legitimate in
context~\cite{koc2026clawhubsignals}. What they measure, though, is agreement
between third-party tools with incommensurable outputs, and they publish no
per-listing cross-tabulation. Nöther et al. name the quadrant directly,
reporting that most of 473 popular skills have some configuration that executes
a malicious action~\cite{nother2026contra}; that is exposure under adversarial
configuration search, which is a different quantity from exposure at default
installation. To our knowledge, prior agent-skill studies have not separately operationalized malicious intent and capability exposure over the same marketplace population or reported their joint distribution.

\subsection{From Static Analysis to Detonation}

Detonating marketplace artifacts is not new and did not begin with agents.
SkillExplorer executed 30{,}801 voice-assistant skills in 2020, driving each with
a synthesized natural-language stimulus and comparing observed data requests
against declared policy~\cite{guo2020skillexplorer}. Hulk exercised 48{,}332
browser extensions six years earlier, driving each with dynamically adapting
HoneyPages and a fuzzer over the event handlers the extension
registers~\cite{kapravelos2014hulk}. In the agent era, Liu et al. detonated 4{,}287
skills in instrumented containers with honeypot credentials and syscall
tracing~\cite{liu2026donotmention}, and SkillScope instantiated 16{,}743 user
tasks over skills drawn from the same two marketplaces this paper
uses~\cite{wu2026skillscope}. We make no claim to be first at this, or largest.

Both agent-era studies apply dynamic execution only after an earlier analysis stage has selected candidate listings. Liu et al.~\cite{liu2026donotmention} collected 98,380 skills, selected 4,287 suspicious candidates using static triage, and dynamically verified those candidates, which corresponds to 4.4\% of their corpus. SkillScope similarly applies replay-based validation to candidates identified through its graph-based analysis \cite{wu2026skillscope}. By contrast, every listing in our sampled cohort was selected for detonation without using a prior static threat or risk signal to determine which listings were executed. This distinction matters because a threat-oriented prefilter may exclude apparently benign listings that nonetheless carry consequential capabilities. Furthermore, this unconditioned selection allows us to estimate exposure prevalence within the sampled cohort and identify high-exposure listings carrying benign verdicts. Because the corpus is popularity-banded rather than a complete census or probability-representative sample, these estimates should not be interpreted as marketplace-wide base rates. Unconditioned execution is therefore a prerequisite for the exposure analysis within this cohort, rather than an independent methodological virtue.

\subsection{Capability Declaration}

Comparing what an artifact declares against what it does is a long line of work.
Stowaway measured over-request in Android
applications~\cite{felt2011stowaway}, Bartel et al. named closing that gap
attack-surface reduction~\cite{bartel2012attacksurface}, and the same comparison
has been carried out for browser extensions~\cite{kapravelos2014hulk} and for
Alexa skills at 199{,}295 artifacts~\cite{edu2021skillvet}. Every one of those
ecosystems had a permission vocabulary, and in every one the finding was that
developers ask for more than they need.

Agent skills invert the polarity, and others argued this before we did. Four separate proposals in the standard's own repository ask for
frontmatter fields to declare environment access, credentials, capabilities and
required permissions~\cite{agentskills2026proposals}; the same absence is
asserted for the sibling protocol in a standards
proposal~\cite{mcp2026sep2385}; and a permission-centric framework for skills
proposes the missing vocabulary outright~\cite{skillguard2026permissioncentric}.
Parts of it are also already measured. Gao et al. report how often the one
permission-shaped field appears across 41{,}662
skills~\cite{gao2026registrytorepository}; an ecosystem scan of 17{,}065 skills
compares declared shell grants against observed shell
use~\cite{skillock2026scan}; and environment and credential access have been
quantified both statically, at 127 listings harvesting environment variables and
52 reaching credentials in a categorized sample of
1{,}218~\cite{liu2026agentskillswildempirical}, and again over a stratified
corpus~\cite{skillguard2026permissioncentric}, and in a sandbox provisioned with
mock credentials~\cite{chen2026credentialsleakedllmagent}.

Agent skills also invert the classic result. In every ecosystem above, the
dominant error was over-request. Here the dominant error is the opposite:
among listings that declare anything, more understate what their bundle can do
than overstate it. The inversion has a mechanical cause. Those ecosystems enforced their permissions, so an application could not use what it had not requested; the skill field restricts nothing, so a bundle can reach far more than it lists. Whether declaring predicts declaring accurately turns out to
be the sharper question, and the two marketplaces answer it in opposite
directions.

\subsection{Framework Mapping}

Findings are mapped to three catalogs: MITRE ATLAS, the OWASP Agentic Top 10,
and the recently published OWASP Agentic Skills Top
10~\cite{owasp2026ast10}. The last of these is the least mature, and its
repository and whitepaper have diverged, so we pin the version mapped against
and report coverage figures only. It matters here for a
specific reason: several of its mitigations presuppose a permission manifest a
reviewer can inspect, which is precisely the artifact
Section~\ref{sec:attack-surface} measures.

\subsection{Contributions}

\begin{itemize}
    \item An exposure-oriented, two-phase attack-surface profiling
    methodology combining whole-bundle static analysis, dependency
    provenance checking, instrumented sandbox execution, and separate model-based assessment of the resulting traces.

    \item An application of the methodology to 6,309 marketplace
    listings selected for execution without prior threat- or
    risk-based filtering, enabling threat and exposure to be
    analyzed independently over the same cohort.

    \item A joint threat--exposure analysis showing that substantial
    capability exposure frequently exists without measured malicious
    intent.

    \item A measurement of declaration gaps, including adoption,
    expressiveness, and agreement between author pre-approvals and
    inferred bundle capabilities.

    \item An empirical comparison showing that static and runtime
    evidence identify materially different aspects of skill risk.

    \item An empirical mapping of threat and exposure findings to MITRE ATLAS, the OWASP Agentic Top 10, and the OWASP Agentic Skills Top 10, showing that the two measurement axes concentrate in different framework categories.
\end{itemize}

\section{Scope and Methodology}
\label{sec:scope-method}

\subsection{Scope}
\label{sec:scope}

The study covers 6{,}309 skill listings published by 2{,}392 distinct
publishers across two marketplaces, 3{,}947 from skillsmp and 2{,}362 from
ClawHub. The two gate publication differently, which
Section~\ref{sec:risk} takes up. All were scanned between 28 July and 25 August
2026, and no report was missing or unreadable.

Selection was constrained by the cost of scanning every listing and by a
preference for the middle of the popularity distribution. Listings on skillsmp
expose a star count, and repositories were sampled by star range into three
bands holding 1{,}533, 1{,}180 and 1{,}234 listings from the highest to the
lowest. ClawHub exposes download counts, and its listings fall into
three download bands holding 1{,}272, 517 and 500, and a residual 73 for which
no download figure resolved, the four together accounting for its 2{,}362
listings. Both sets of bands describe how the corpus was drawn and not
how either marketplace is distributed. The extreme popularity head was
excluded
deliberately, since those listings are the ones already under scrutiny, and so
was the unused tail, since a skill nobody installs exposes nobody.

Publisher concentration is uneven and every per-marketplace rate should be read
against it. A publisher contributes 2.6 listings on average and one at the
median, 19 publishers contribute more than twenty, and the largest single
publisher contributes 266. Concentration also differs between the two: a
publisher contributes 4.1 listings on average on skillsmp against 1.7 on
ClawHub, so rates on skillsmp in particular are closer to rates per publisher
than per skill.

Two limits on generalisation follow. The corpus is a sample and not a complete
pass over either marketplace: contemporaneous crawls put one of them alone at
15{,}082 listings in December 2025~\cite{liu2026agentskillswildempirical} and
73{,}193 a month later~\cite{liu2026donotmention}. Popularity is also not
measured the same way on both, since skillsmp exposes stars and ClawHub exposes
download counts, so any analysis by popularity runs separately
on each against a different variable, and both are point-in-time scrapes that
move daily.

\subsection{Methodology}
\label{sec:method}
Our methodology combines whole-bundle static inspection with instrumented sandbox execution to measure threat, exposure, and capabilities separately. Automated findings are qualified through sampled human review and targeted re-scans. Fig.~\ref{fig:profiling_pipeline} summarizes these analytical stages; the following subsections describe their implementation and limitations. 

\begin{figure}[H]
\centering
\fbox{%
\begin{minipage}{\dimexpr\columnwidth-2\fboxsep-2\fboxrule\relax}
\small
\textbf{Input:} A sampled skill listing and its complete bundle.

\begin{enumerate}
\item \textbf{Inspect the bundle.}
Read frontmatter, instructions, scripts, and bundled resources
with the static analyst. Supplement model findings with checks
for hard-coded secrets, piped installers, invisible Unicode,
and opaque binaries.

\item \textbf{Check referenced dependencies.}
Where dependencies are declared, check whether referenced
packages, repositories, and accounts resolve, whether versions
are pinned, and whether hosts are takeover-prone. Resolution
does not establish authenticity or intent.

\item \textbf{Exercise the skill.}
Select listings for sandbox execution without prior threat-
or exposure-based filtering. Record processes, tool calls,
and network activity, together with execution status.

\item \textbf{Assess runtime evidence.}
Where a judge is configured, use a model separate from the
static analyst to assess the recorded trace. Retain missing
or unavailable assessments as such.

\item \textbf{Report distinct measurements.}
Record the threat verdict, exposure score and level, inferred
capabilities, and amplification factors. Keep static and
runtime evidence distinguishable. Apply the documented
crosswalk and cohort-level executor filtering when producing
aggregate tables.

\item \textbf{Qualify the findings.}
Use the human-review samples and re-scans to assess selected
outputs and report the resulting validation and coverage limits.
\end{enumerate}

\textbf{Output:} Listing-level measurements and aggregate
findings, interpreted under the limitations in Section III-B6.
\end{minipage}}
\caption{Overview of the disclosed profiling and validation
stages. The numbering organizes the analytical operations;
it does not specify production scheduling or reproduce the
proprietary prompts, scoring rubric, or orchestration.}
\label{fig:profiling_pipeline}
\end{figure}

\subsubsection{Two tiers}
\label{sec:method-tiers}

Every listing passes through an automated two-phase scan. A second, human tier
adjudicates samples of that output under written evidentiary rules, and no
flagged listing is reported individually before it has been adjudicated. That
tier's design is given here, because the review is
partial and its coverage bounds what the verdicts support.

The human cohort covers 410 listings, of which 91 carry a flagged verdict and
319 do not. Flagged verdicts are reviewed under the same rules; unflagged listings are sampled. Two limits bound that cohort. It is drawn from one marketplace, so it says nothing about the flagged
population of the other. And the reviews were in progress at the time of
measurement, so no precision figure is reported here and none should be read
into the cohort's existence.

Recall on the benign majority is a separate measurement and it is complete. 140
listings drawn at random from the whole benign population were read in full,
every file in every bundle, without filename triage. None was a missed threat,
which places a one-sided 95\% upper bound of approximately 2.1\% on the benign miss rate. That
bound is not a precision figure and does not substitute for one.

Recall reviews ran as parallel independent readers with no visibility into each
other's conclusions, each recording the files read, a verdict, and the lines that
decided it. Every claimed missed threat and every borderline case is re-verified
against the bundle bytes by the lead reviewer before entering any published
number. That step rejected two plausible reviewer claims: a device fingerprint that proved to be a locally
generated random identifier, and a self-granted administrative session that
crossed no boundary because the script already ran with those privileges.

Verdict changes are treated asymmetrically. An increase in severity applies
automatically, because raising is the safe direction. Every decrease is withheld
for human review, because downgrading a flagged listing is the direction that
can conceal a true positive. An increase driven only by a pattern already known
to produce false positives is also withheld.

\subsubsection{What is measured}
\label{sec:method-quantities}

Three quantities are recorded per listing, and they are deliberately distinct.

\textbf{Threat} is malicious intent, carried by the published verdict. The
verdict is a thresholded score: malicious at or above 0.70 and suspicious at or
above 0.50.

\textbf{Risk} is the severity of a listing's exposure, taken from an agentic
exposure score on a scale of zero to ten~\cite{owasp2026aivss}, from which a
level of low, medium, high or critical is banded.

\textbf{Attack surface} is the descriptive capability footprint over which the risk judgment is made: what the bundle contains evidence of before any severity or intent is assigned. Capabilities are inferred by the static phase from reading the bundle, and are reported as eleven groups collapsed from 70 model-emitted labels. This vocabulary is close to the categories used in package capability analysis~\cite{ferreira2021npm,google2023capslock}, with a credential class added and filesystem read separated from write. An agentic amplification component records factors such as tool use, autonomy, memory, and self-modification~\cite{owasp2026aars}.

The two judgment axes are operationalized and scored separately by construction. Each finding carries an axis label and is counted on only one axis. The threat verdict derives from a threat score, whereas the exposure level derives from a separate exposure scale. The capability footprint does not contribute to the threat score; instead, it describes the authority available to a skill irrespective of whether malicious intent is present. This separation concerns the design of the measurement process and does not assume statistical independence between threat and exposure.

\subsubsection{Static analysis}
\label{sec:method-static}

A model analyst reads the complete bundle, every file including frontmatter, scripts, and bundled resources, against a security rubric, and emits findings on both axes. Two models were used across the corpus, Claude Haiku 4.5 on 5{,}791 listings and Claude Sonnet 4.6 on the remaining 518. A proprietary rubric and prompt template guide the model analyst in evaluating the complete bundle. The production templates and internal orchestration are not released as described under Data Availability. Four deterministic heuristics supplement it, covering signals we wanted reproducible: hardcoded secrets, piped installers,
invisible Unicode and opaque binaries. The static phase produces 4{,}246 findings across the corpus.

Those findings carry free-text titles, 2{,}458 of them distinct, most differing only in phrasing. The pattern tables in Sections~\ref{sec:threat} and~\ref{sec:risk} are produced by collapsing every title onto a named pattern
through an ordered keyword crosswalk. That crosswalk is a presentation layer and not a detector: it groups what the analyst already found and adds nothing. It matches 95\% of static threat findings and 87\% of static risk findings, with the remainder falling to a stated default bucket on each axis, and a stratified sample of 74
assignments was audited by hand. The mapping to published catalogs is separate and mechanical, taken from the codes each finding already carries.

Each static finding carries one scheme code. Of the 4{,}246, 2{,}508 route to an Agentic Skills Top 10 entry and 1{,}710 to an Agentic Top 10 entry, with 28 carrying neither, so the two schemes partition the findings between them. MITRE ATLAS is not comparable on that basis, because a finding may carry several techniques, so the 3{,}789 technique mentions recorded across fifteen techniques are not a count of findings and cannot be put on the same scale. Which scheme a finding routes to is a property of the vocabularies and not a measure of how much of the population is hazardous.

The distinction matters when the pattern tables are read across phases in Section~\ref{sec:static-vs-dynamic}, because both phases enter the same crosswalk from different evidence.

\subsubsection{Detonation}
\label{sec:method-dynamic}

All but three listings were executed. An agent runs the skill in an isolated sandbox, and the processes, tool calls, and network egress it produces are recorded. A separate judge model, not the static analyst, then assesses the resulting trace. Three roles are filled by three configurations. The static analyst is Claude Haiku 4.5 on 5{,}791 listings and Claude Sonnet 4.6 on 518. The detonating agent is Claude Haiku 4.5. The judge is Gemini 3.1 Pro Preview on
5{,}769 listings, with a Claude model on five and none recorded for the cohort that ran without a judge, so the two phases are separated by vendor as well as by prompt and evidence.

Executor activity is separated at aggregation, not per listing. Any process or tool name appearing in at least half of all listings is dropped as executor baseline, and four executor runtimes are dropped by name regardless of how often they appear, which leaves 522 process names that are distinctive to some subset of listings. That is a prevalence correction taken across the cohort, not a subtraction of a listing's own control run.

\subsubsection{Provenance analysis}
\label{sec:method-provenance}

Dependencies referenced by a listing are resolved rather than assumed, as are its source repository and publishing account, and each dependency is checked for version pinning and for a takeover-prone host. The resulting counts are reported
with the supply-chain surface in Section~\ref{sec:attack-surface}.

The phase is legitimately absent for a listing that declares no dependency at all, which is most of the corpus, so its denominator is the declaring population and not the whole.

The check establishes existence, not intent. It confirms that a named dependency, repository, or account resolves; it does not confirm that the one which resolves is the one the author meant. A lookalike package registered by an attacker resolves and is reported as present. No name-similarity detection
runs, so we make no claim about typosquatting here; the only evidence would be unstructured analyst narrative.

\subsubsection{Limits of the measurement}
\label{sec:method-limits}

Five limits follow~\cite{rossow2012prudent}.

Verdicts are thresholded model scores and not findings of fact.
Section~\ref{sec:method-tiers} gives the adjudication that qualifies them.

Scanning is not deterministic. Verdict reproducibility was measured at 84\% over
89 re-scans, with the differences concentrated at the flagged-severity thresholds of listings that were already suspect. Two benign listings did flip to flagged on re-scan, so benign labels are about 98\% self-consistent. Per-listing labels near a threshold carry this uncertainty, and no interval is placed on the aggregate rates reported here. Re-scan disagreement measures instability but does not establish that classification errors are directionally balanced or cancel in aggregate. A further constraint is publisher concentration: one publisher contributes 266 listings, so listings are not independent draws and an interval calculated under an independence assumption would be too narrow.

Runtime observations cover only what the sandbox detonation happened to exercise, which bounds them from below. A recall check found a family of listings whose bundled setup script writes persistent agent configuration but is never referenced from any skill file, so detonation never invokes it.

Most listings ran once with no control. A runtime observation is therefore co-occurrent with a listing and not attributed to it, and executor activity is separated only by prevalence across the cohort. On the smaller cohort that did run a control, shell invocation and environment access appear in the control arm on every listing, so neither separates skill behavior from executor behavior at all.

Capability and amplification labels are model-emitted free text and were canonicalized after the fact, so counts for rare labels are indicative only.

\section{Findings}
\label{sec:findings}

\subsection{High-Level Findings}
\label{sec:findings-highlevel}

Three properties of the same 6{,}309 listings were measured, and they select populations of very different sizes. 4{,}431 listings carry at least one elevated capability. 3{,}089 carry risk, meaning an exposure level at medium or above or a static risk finding. 338 carry a non-benign verdict. Capability is the common condition, exposure is frequent, and malicious intent is rare.

The relation between the last two runs in one direction only. Every listing carrying a threat verdict sits at medium exposure or above, so on this corpus threat implies risk. The
converse fails by a wide margin: 1{,}306 listings sit at high or critical exposure with a benign verdict, against 315 flagged ones, so four in five listings at the top two exposure levels carry no threat verdict. Those benign
labels are not simply undetected malice: a random sample of 140 listings drawn from the whole benign population was read in full by hand, and none was a missed threat, which bounds the benign miss rate approximately 2.1\%
(Section~\ref{sec:method-tiers}). A review process that asks only whether a listing is hostile therefore leaves the larger population unexamined, and that population is what
Sections~\ref{sec:risk} and~\ref{sec:attack-surface} are about.

Section~\ref{sec:threat} finds threat rare and concentrated, with one pattern, offensive tooling published as a skill, holding roughly a third of the listings carrying any threat signal. Section~\ref{sec:attack-surface} finds the permission model close to inert: 770 of the 6{,}061 listings whose frontmatter parses declare anything at all, and 457 of those understate what their bundle can do. Section~\ref{sec:static-vs-dynamic} finds the two analysis phases disagreeing on
most of what either one flags, which is why we report both.

Every figure below is a single-version measurement from one scan of each listing, and Section~\ref{sec:method-limits} states the reproducibility bounds that qualify them.

\subsection{Threat Profiling}
\label{sec:threat}

\subsubsection{How much of the corpus carries a threat}

338 listings carry a non-benign verdict, 206 judged malicious and 132 suspicious (Table~\ref{tab:joint}), which is 5.4\% of the corpus. Two adjacent figures answer slightly different questions. Widening from the thresholded verdict to any threat-axis finding or any threat score above the
lower threshold gives 345 listings. The pattern counts below sum differently again, because one listing can match several patterns. We therefore use the verdict-based estimate as the primary prevalence measure, while the adjacent measures capture broader threat signals.

That rate is close to what static analysis with model assistance has previously reported for this ecosystem, where roughly one listing in twenty showed likely malicious intent across 31{,}132 skills~\cite{liu2026agentskillswildempirical}. Two methods with little in common arriving at a similar rate is weak evidence, but it is evidence, and it is the only external calibration available for a figure of this kind.

\subsubsection{What the threats are}

\begin{figure}[t]
\centering
\includegraphics[width=\columnwidth]{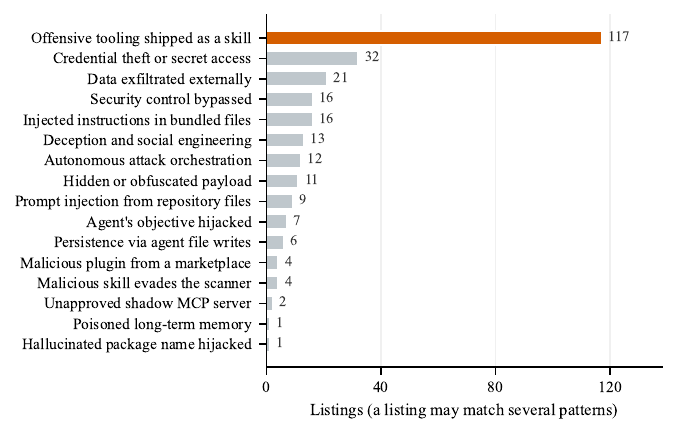}
\caption{Threat patterns across the corpus. One pattern holds more listings than the next four combined, and it is dual-use tooling rather than covert malware. A listing may match several patterns, so the column does not sum to the threat
population.}
\label{fig:threatpat}
\end{figure}

Threat is rare and concentrated. Fig.~\ref{fig:threatpat} gives the distribution across sixteen patterns. Measured against the 345 listings carrying any threat signal, the first row accounts for roughly a third on its own, and it holds more listings than the next four patterns combined. 

The dominant category is offensive tooling packaged and published as a skill: material that is dual-use by construction. Whether it belongs on a marketplace is a policy question. Strip it out, and the remainder of the threat surface is a long, thin tail in which no pattern
reaches forty listings and most reach fewer than fifteen. A listing may match more than one pattern, so the column does not sum to the threat population.

Read against Section~\ref{sec:attack-surface}, that is the section's central result. Seven listings in ten carry elevated capability and one in twenty carries a threat, so the two populations are of entirely different sizes.

Severity is recorded per finding, not per listing, and the flagged population is severe when it is flagged at all: of 383 threat findings, 186 are critical and 167 high, leaving 30 across the three lower levels combined. Findings are concentrated at the top of the severity scale.

\subsubsection{Standards mappings}

Threat findings map to three published catalogs, and the mapping unit is findings, not listings throughout. Against the Agentic Skills Top 10, 240 findings fall under malicious skills, with supply-chain compromise at 12, poor scanning at 11, insecure metadata at 9, and untrusted external instructions at 7. Against the Agentic Top 10, identity and credential abuse leads at 38 findings, followed by goal and instruction hijack at 27 and tool misuse at 23. Against
MITRE ATLAS, external harms accounts for 215, unsecured credentials for 55, prompt injection for 38 and supply-chain compromise for 37.

The three schemes disagree about how concentrated the threat population is, and the disagreement is informative. Under the skills-specific catalog, a single entry absorbs 240 of the 280 threat-axis findings it maps, because that catalog has a category named for exactly this population. The two general-purpose schemes spread the same findings much wider, the leading entry taking under two fifths of the agentic mappings and about half of the ATLAS ones. A taxonomy built for the
broader field does not resolve this population into one bucket, and the one built for skills does, which says more about the catalogs than about the threats. An ATLAS row counts findings carrying that technique, and a finding may carry several, so those rows sum past the finding total.

\subsubsection{Threat by marketplace and popularity}

The two marketplaces differ, and not in the direction the difference in their install-time controls would suggest. Measured on the broader threat signal rather than the published verdict, skillsmp carries a rate of 6.6\%, with
260 listings of 3{,}947, against 3.6\% on ClawHub, where 85 of 2{,}362 carry a threat. We report this without claiming it
measures either marketplace's security program. The two cohorts were sampled differently and on different popularity variables, and ClawHub's rate in particular is sensitive to that: its coverage grew from 862 listings to 2{,}362 when download-banded cohorts were added, and the rate moved with them. The comparison supports only one thing: no gating policy is visible in these rates. It does not rank the marketplaces.

Popularity does not screen threat either, and the two marketplaces disagree about which way it points. On skillsmp, where popularity is a star count, the rate runs 5.8\%, 7.4\% and 6.6\% from the lowest band to the highest, with
no trend. On ClawHub, where popularity is a download count, it runs 4.6\%, 2.9\% and 2.0\% from the lowest band to the highest. Neither adjacent step separates, and only the two ends do, so the ClawHub series is a difference between its extremes and not a trend across its bands. Both sets of bands rest on the same broader basis as the marketplace rates above. Neither pattern supports treating popularity as a safety signal, and the contrast with Section~\ref{sec:risk} is that exposure tracks popularity where malice
does not.

\subsection{Risk Profiling}
\label{sec:risk}

\subsubsection{How much of the corpus carries risk}

A listing counts as carrying risk when its exposure level reaches medium or above, or when the static phase records at least one risk-axis finding against it. 3{,}089 listings meet that test, which is 49.0\% of the corpus. Fig.~\ref{fig:exposure} plots the stricter subset, the 1{,}621 listings whose exposure score reaches seven or above.

Set beside the 5.4\% carrying a threat verdict, the two axes select populations of very different sizes. The two are separately thresholded, and their ratio is not itself a quantity, but the gap is the reason they are worth reporting apart.

The exposure score behind the level is bimodal, and its shape matters more than its average. 3{,}128 listings score zero: prose guidance, reference material, and style guides carrying nothing the exposure scale can rate. Among the 3{,}181 that do
score, the mean is 6.81 and the median 7.1, and 1{,}621 reach seven or above with 749 of those in the critical band. A corpus-wide mean of 3.44 therefore describes neither population.

Those figures need reconciling with Section~\ref{sec:attack-surface}, which reports 4{,}431 listings carrying an elevated capability. Since 4{,}431 and 3{,}128 exceed the corpus between them, at least 1{,}250 listings carry an elevated capability and still score zero exposure. Capability presence and exposure magnitude are different measurements, and a listing with one narrow capability and no amplification scores at the bottom of the scale without being inert.

\subsubsection{Threat and risk on the same listings}

The value of reporting the two axes separately lies in the extent to which they disagree. Table~\ref{tab:joint}
crosses the published verdict against the exposure level for every listing, and the disagreement is structural.

\begin{table}[t]
\caption{Verdict against exposure level, all 6{,}309 listings. Computed from the scan index. The row totals are the published verdict distribution and the column totals the published risk-level distribution, both reproduced exactly.}
\label{tab:joint}
\centering\small
\begin{tabular}{lrrrrr}
\toprule
 & Low & Medium & High & Critical & Total \\
\midrule
Benign      & 3{,}658 & 1{,}007 & 745 & 561 & 5{,}971 \\
Suspicious  &       0 &      23 &  75 &  34 &    132 \\
Malicious   &       0 &       0 &  52 & 154 &    206 \\
\midrule
Total       & 3{,}658 & 1{,}030 & 872 & 749 & 6{,}309 \\
\bottomrule
\end{tabular}
\end{table}

Neither of the results that follow is visible in the marginals. No listing carrying a threat verdict sits in the low exposure band, so on this corpus a threat implies exposure at medium or above. The converse does not hold: 1{,}306 benign listings sit at high or critical exposure, against 315 flagged ones. Among listings at the top two exposure levels, four in five carry no threat verdict at all.

That population is what this paper measures: it is not detectable as malice; it is
larger than the malicious population by a factor of four, and a review process tuned to intent will not see it.

\subsubsection{What the risks are}

\begin{figure}[t]
\centering
\includegraphics[width=\columnwidth]{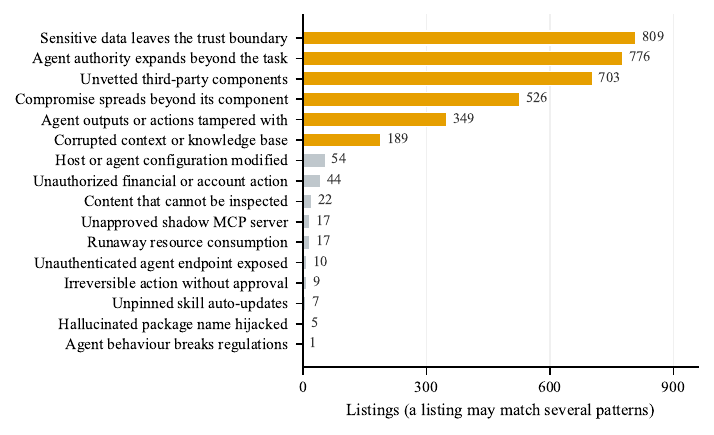}
\caption{Risk patterns across the corpus. Compared with the threat distribution,
risk spreads across many patterns. A listing may
match several patterns.}
\label{fig:riskpat}
\end{figure}

Where threat is concentrated into one pattern, risk spreads.
Fig.~\ref{fig:riskpat} shows six patterns above 180 listings and a tail below sixty, with the leading three separated by 106 listings. Sensitive data crossing
a trust boundary leads at 809 listings, agent authority expanding past the task follows at 776, and unvetted third-party components at 703. These counts are static-phase findings, and a listing may match several patterns, so the column neither sums to nor is a share of the 3{,}089 above.

The leading row needs care, because it absorbs the lethal trifecta, meaning private data reachable alongside untrusted content and an outbound channel. That composition holds 628 listings on its own, and it names an available exfiltration path. No leakage was observed. Separately, the row for content that could not be inspected, at 22 listings, reports a property of the artifact rather than a failure of the scan.

\subsubsection{Severity, and why two ladders disagree}

Risk severity is recorded per finding: 2 findings at critical, 1{,}262 at high, 915 at medium, 1{,}487 at low and 197 informational. Against an exposure distribution that puts 749 listings in the critical band, those two figures look
irreconcilable. A finding severity rates one observation about a bundle. An exposure level is banded from a score aggregating how far a listing's capability reaches together with its amplification factors. A listing therefore reaches the critical band on the breadth of what it can touch while no single observation about it is individually critical. The two measures operate at different levels and therefore do not represent a scoring inconsistency.

\subsubsection{Standards mappings}

Against the Agentic Skills Top 10, 1{,}051 risk findings fall under over-privileged skills and 799 under supply-chain compromise, with untrusted external instructions at 202 and weak isolation at 120. Against the Agentic Top 10, goal and instruction hijack leads at 752, tool misuse at 376, and identity and credential abuse at 212.

The two axes concentrate in different places. Over-privileged skills and supply-chain compromise, the two entries that barely registered on the threat axis, take over four fifths of
the risk findings mapped under the Agentic Skills Top 10. Neither describes an attacker; both describe an install-time property of a bundle that a reviewer could inspect before anything runs.

\subsubsection{Risk by marketplace and popularity}

\begin{figure}[t]
\centering
\includegraphics[width=\columnwidth]{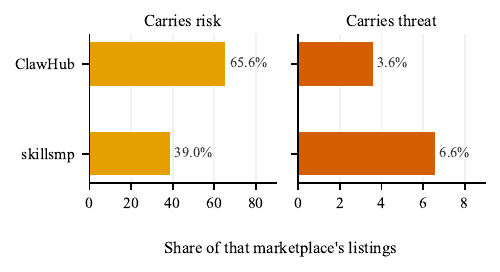}
\caption{The two axes order the marketplaces differently. ClawHub carries risk on 65.6\% of its listings against 39.0\% on skillsmp, while its threat rate, measured on the broader signal of Section~\ref{sec:threat}, is 3.6\% against
6.6\%. Which marketplace a single-axis study ranks higher depends on the axis chosen, though only the risk leg of that ordering has held across corpus revisions. The panels carry independent scales, because the threat rates are an
order of magnitude smaller and the reversal is not visible beside the risk bars.}
\label{fig:mkt}
\end{figure}

The two marketplaces are ordered differently by the two axes.
Fig.~\ref{fig:mkt} sets the orderings side by side: ClawHub carries
risk on 65.6\% of its listings against 39.0\% on skillsmp, while on the
threat axis, measured on the broader signal used in
Section~\ref{sec:threat}, the ordering reverses at 3.6\% against 6.6\%.

The risk leg has held across corpus revisions.
The threat leg has not: ClawHub's threat rate fell from 6.3\% on the 862 listings covered before the download-banded cohorts to 3.6\% across all 2{,}362, so the two marketplaces sat within a third of a percentage point of each other on this axis before the addition, and the ordering survived it while the size of the gap did not. The current corpus covers ClawHub more completely, so it is the better estimate, but the threat ordering rests on a sample that recently moved, and the risk ordering does not.

The risk indicator and a documentation filter
are close relatives on this corpus, since 3{,}181 listings carry any exposure at all and 3{,}089 carry risk. A marketplace whose catalog holds proportionally less inert prose will therefore show a higher risk share for reasons that have nothing to do with its review policy. We report the marketplace comparison for completeness and draw no conclusion from it about either marketplace's security
program. Per its own published documentation, skillsmp applies
automated security scanning at install time, while ClawHub describes a publish-and-download flow with no documented mandatory pre-publication review.
That is context; we did not evaluate either control.

Popularity behaves consistently here, unlike on the threat axis. The share carrying risk rises with popularity on both marketplaces: from 35.5\% through 38.0\% to 42.6\% across skillsmp's star bands, and from 62.4\%
through 68.5\% to 69.6\% across ClawHub's download bands. More
widely adopted listings do more, and doing more is what the risk axis measures. Popularity screens neither axis, and on this one it points the wrong way mildly.

\subsection{Attack Surface Profiling}
\label{sec:attack-surface}

\subsubsection{Inferred capability footprint}

\begin{figure}[t]
\centering
\includegraphics[width=\columnwidth]{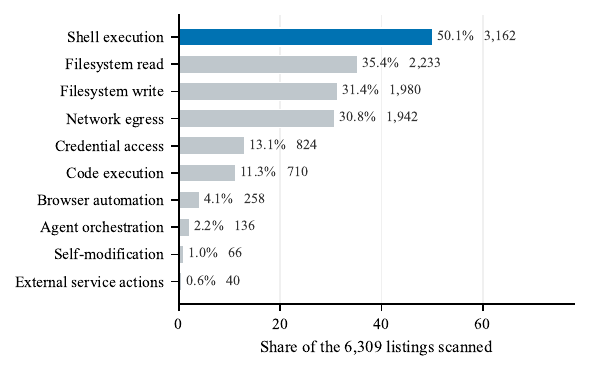}
\caption{Capability footprint of the corpus, grouped from 70 model-emitted
labels. A residual group of 25 listings carrying only uncanonicalized labels is
omitted.}
\label{fig:caps}
\end{figure}

Capability here is descriptive and says nothing about intent: it is an action for which the bundle contains supporting evidence, irrespective of severity or intent. 4{,}431 listings, seven in ten, carry
at least one elevated capability, and a listing carries 1.9 capabilities and 1.6 amplification factors on average. Fig.~\ref{fig:caps} gives the distribution, grouped from 70 model-emitted labels.

Half the corpus carries inferred shell-execution capability. That single figure bounds most of
what follows, because a shell is a general capability and the classes beneath it are largely reachable through it.

397 listings hold both credential access and network egress, which forms a potential path for data to leave the host. 628 satisfy the lethal trifecta of access to private data, exposure to untrusted content and an outbound channel. 1{,}621 score at or above seven on
the exposure scale. Two smaller rows are lower bounds derived from finding titles rather than from capability: 17 listings were flagged for a destructive operation, and 46 for a persistence mechanism, and any listing with a shell could
attempt either.

\subsubsection{Agentic amplification}

Amplification records the properties that make a capability reach further in an agent than it would in a script. Tool use is near-universal at 4{,}087 listings,
non-determinism appears in 1{,}814, autonomy in 1{,}501 and memory use in 1{,}229. The three rarer factors bear on blast radius: dynamic identity in 560 listings, multi-agent interaction in 469, self-modification in 180.

\subsubsection{Observed runtime activity}
Static capability is inferred from the bundle; detonation records what occurred during one session.
The most frequently recorded tool is file writing at 2{,}571 listings, followed
by invoking another skill at 1{,}479 and editing at 617. Shell invocation appears
for 518, environment reads for 518, credential reads for 344, and outbound requests for 286. At the process level, the 522 process names that survive the prevalence filter of Section~\ref{sec:method-dynamic} are led by file and text
utilities at 1{,}819 listings, language runtime at 1{,}448 and system introspection at 1{,}427. The groups that bear on security sit below those: package managers and installers at 1{,}105, network clients at 632, privilege and system modification at 617, and skill-bundled or fetched binaries at 480.

The gap between the two views is large and is not a contradiction. Half the corpus can reach a shell, and roughly one in twelve was observed to use one, because a single sandbox session exercises few of a bundle's code paths.
Runtime attribution is therefore a lower bound throughout, and the static comparison carries the section that follows.

\subsubsection{Supply-chain surface}

2{,}871 listings declare a dependency of some kind, and 2{,}830 of those reference at least one that is not pinned to a version. Read against the
declaring population, not the corpus, that is 99\%: pinning is the exception and not the norm. Among them, 75 resolve a dependency to a host that
is takeover-prone. Together, this is the surface through which a listing
that is benign today becomes something else tomorrow without its author acting.

\subsubsection{Declared against actual capability}

\begin{figure}[t]
\centering
\includegraphics[width=\columnwidth]{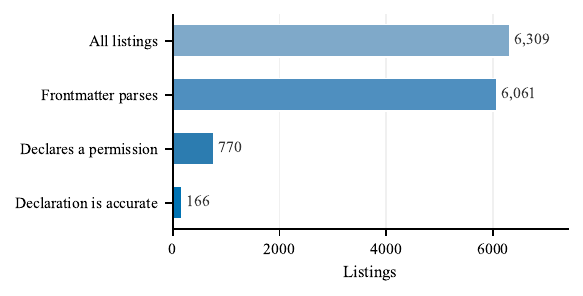}
\caption{The pre-approval field in use. Almost every listing carries parseable frontmatter, one in eight pre-approves tools in it, and fewer than a quarter of those lists match what the bundle can do. The field grants, and does not restrict, so a mismatch in either direction is a finding.}
\label{fig:decl}
\end{figure}

The format lets an author list tools the skill pre-approves. Of the 6{,}061 listings whose frontmatter parses, 770 declare anything at all, which is 12.7\%. That is the section's first finding, and it precedes every question about accuracy: absent a declaration, there is nothing to check a bundle against, and no statement of what the author intends the skill to reach. Fig.~\ref{fig:decl} shows the narrowing.

Among the 770 that do declare, the comparison runs over the five classes both vocabularies can express. Credential and environment access are excluded because nothing in the format can be compared against them. The format carries an
optional free-text compatibility string whose stated purpose is to indicate environment requirements, and the specification contemplates network access being mentioned inside it. Prose cannot be matched against an observed
capability, so there is nothing an analyzer can check either class against, and including them would make every listing look under-declared by construction. Table~\ref{tab:declared} gives the result.

\begin{table}[t]
\caption{Declaration accuracy among the 770 declaring listings. The first two rows overlap on the 157 listings that under-declare one class while over-declaring another, so the percentage column sums past 100. The disjoint
block below partitions the same 770.}
\label{tab:declared}
\centering\small
\begin{tabular}{lrr}
\toprule
 & Listings & \% \\
\midrule
Under-declared & 457 & 59 \\
Over-declared  & 304 & 39 \\
Aligned        & 166 & 22 \\
\midrule
\multicolumn{3}{l}{\emph{Disjointly}} \\
Aligned only     & 166 & \\
Under only       & 300 & \\
Over only        & 147 & \\
Both directions  & 157 & \\
\bottomrule
\end{tabular}
\end{table}

The dominant error is understatement, which inverts the result established for mobile applications, browser extensions and voice assistants, where the characteristic failure was asking for more than was needed. Here a declaration is more likely to understate a bundle than to overstate it. The two directions fail differently. An understated declaration tells a reviewer the listing is narrower than it is, but leaves the agent's confirmation prompts in place for what it omits. An overstated one removes prompts for tools the bundle does not need, and is the one direction in which a declaration changes what the session permits rather than only misdescribing it; 304 of the 770 declaring listings do so.

\begin{figure}[t]
\centering
\includegraphics[width=\columnwidth]{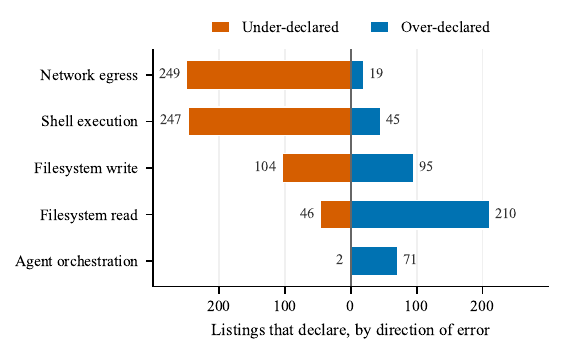}
\caption{Direction of declaration error, by capability class, among the 770 listings that declare. Shell execution and network egress are understated; filesystem read is overstated by a comparable margin, and agent orchestration in the same direction at a smaller scale.}
\label{fig:misdec}
\end{figure} 

Which class is misdeclared, and in which direction is a property of the vocabulary and not the authors, and the two directions separate cleanly by class (Fig.~\ref{fig:misdec}). Shell execution and network egress are understated by a
wide margin. Filesystem read is overstated by a comparable one, and agent orchestration in the same direction at a smaller scale. The classes an author can name concretely get over-requested in the familiar way; the classes that are diffuse, a shell or an outbound connection, get missed.

Declaring often and declaring accurately come apart, and the two marketplaces separate cleanly on it. ClawHub declares roughly twice as often as skillsmp, with 388 of its 2{,}196 parsing listings against 382 of 3{,}865, which is 17.7\% against 9.9\%. Yet 88\% of its declaring listings
understate their bundle against 31\% on skillsmp.

Runtime evidence corroborates the direction without carrying it. Of the declaring listings, 91 were recorded exercising at least one comparable capability, and 54 of those exercised a class they had not declared.

The per-marketplace comparison for both axes, together with the sampling and review-posture caveats that qualify it, is reported in Sections~\ref{sec:threat} and~\ref{sec:risk}.

\subsection{Static Analysis Against Detonation}
\label{sec:static-vs-dynamic}

Section~\ref{sec:why-new} argued that execution has to be unconditioned for the risk axis to have a base rate at all. That leaves a separate question open, and this subsection answers it: how much of each axis either phase would have missed on its own.

\subsubsection{How often the phases agree}

Table~\ref{tab:phase} crosses the phases against each other. Threat thresholds each phase's own score at one half and risk uses the presence of at least one finding from that phase, so neither column is comparable to the published verdict counts, and three listings carry no threat score from one phase and fall outside that column entirely.

A detonation-only flag needs reading against the caveat in
Section~\ref{sec:method-limits}, which holds that a runtime observation is co-occurrent with a listing and not attributed to it. The caveat governs capability attribution, meaning whether a given behavior originated in the skill or in the executor. A flag here is a judge's finding anchored to that listing's own trace, and the judge assessed no other listing's
trace, so the finding is located even where the behavior is not separable from the executor. What the caveat costs is the converse: a listing the executor would have flagged anyway cannot be distinguished from one the skill drove. Two
effects therefore work against each other in this column, since a single session also exercises few of a bundle's code paths, and the count is a localized observation and not a clean bound in either direction.

\begin{table}[t]
\caption{Listings flagged by each phase. Threat thresholds each phase's own score at one half; risk uses at least one finding from that phase, from either phase rather than the static-only test of Section~\ref{sec:risk}.}
\label{tab:phase}
\centering\small
\begin{tabular}{lrr}
\toprule
Flagged by & Threat & Risk \\
\midrule
Both phases     &  11 &   978 \\
Static only     & 292 & 1{,}653 \\
Detonation only &  33 &   708 \\
\midrule
Either phase    & 336 & 3{,}339 \\
\bottomrule
\end{tabular}
\end{table}

On the threat axis, either phase flags 336 listings and both flag 11. Static analysis accounts for 303 of them and detonation for 44, so the phases agree on one listing in four of what detonation flags and one in twenty-eight of what
static analysis flags. Detonation's contribution here is small and heavily qualified: it reports only three threat patterns at all, two of which are already known to produce false positives, and 9 of its 33 exclusive listings fall in the human review cohort.

The two phases agree more often on the risk axis and still disagree on most of the listings either one flags. 978 listings are found by both, 1{,}653 by static analysis alone and 708 by detonation alone, so of the 3{,}339 either phase
flags, 2{,}361 are visible to one phase only. That total runs 250 above the 3{,}089 reported in Section~\ref{sec:risk}, and the two are counting different things: the earlier figure combines exposure level with static findings, while this column admits a detonation finding on a listing whose exposure level sits below medium.

\subsubsection{The phases see different kinds of things}

The two phases can observe different things. The argument is from what each phase does; the counts below illustrate it. Detonation watches data move and
watches an action reach past the component that took it, which is behavior and has to be run to be seen. Static analysis reads what a bundle declares it will pull in and how far its remit is written, which is text and does not need to
run. A listing that imports an unpinned dependency and never exercises it in a single session is invisible to detonation and plain to static analysis, and a listing that reaches a destination only under a particular input is the reverse.

The pattern counts follow that prediction in both directions. Detonation reports compromise spreading beyond its component on 730 listings against 526 from static analysis, while static analysis reports agent authority expanding beyond the task on 776 against 182, and unvetted third-party components on 703 against one. Both phases are classified through the same taxonomy crosswalk, but their findings enter it from different evidence, so these counts establish the direction of each gap and not its size.

One further comparison points the same way and needs a caveat the others do not. Detonation reports sensitive data crossing a trust boundary on 1{,}066 listings against 809, but the static row is dominated by the exfiltration-path
composition described in Section~\ref{sec:risk}, where nothing was observed moving, so the two counts are not like-for-like.

These gaps do not establish that either phase is more sensitive than the other, and they are not evidence that both phases detect one underlying property with independent errors. A common-target reading would have to explain why the
direction of the gap reverses by pattern, which independent errors on a shared target do not produce.

Elsewhere this paper treats runtime observation as a lower bound, because a single session exercises few of a bundle's code paths. That bound applies to capability attribution, meaning what a listing was seen to exercise. The counts
here are risk findings, not capability attributions, and the two quantities should not be read against each other.

\section{Implications for marketplace review}

The declaration gap also suggests a more useful role for capability descriptions. Rather than treating author-written frontmatter as a verified account of a skill, a marketplace or independent assessor could derive a profile from the submitted bundle, identifying inferred capabilities, supporting evidence, and uncertainty. The profile could be bound to the version of the bundle inspected and attested by its assessor, so that a reviewer can verify who issued it and which artifact it describes. Existing work on automated skill permission manifests informs profile generation~\cite{skillguard2026permissioncentric}, while software attestation specifications provide a model for binding statements to artifacts~\cite{in_toto_attestation}. Neither a signature nor a generated profile would guarantee that every possible behavior was discovered: static inference can be wrong, and one sandbox run exercises only some code paths. These findings motivate independently generated profiles; this study does not implement or evaluate a profile-generation or attestation system. Fig.~\ref{fig:illustrative_profile} illustrates how such a profile could present capabilities, supporting evidence, and review questions. The example is hypothetical and is not a listing or scan result from the corpus.

\begin{figure}[t]
\centering
\fbox{%
\begin{minipage}{\dimexpr\columnwidth-2\fboxsep-2\fboxrule\relax}
\small
\textbf{Illustrative capability profile}\\
\textit{Hypothetical report-upload skill; not a corpus result.}

\medskip
\textbf{Bundle scenario.}
The instructions invoke a bundled Python script. The script
reads a local report, retrieves an API token from an environment
variable, and sends the report to an external service.

\medskip
\textbf{Capabilities supported by the hypothetical bundle:}
\begin{itemize}
\item \textbf{Shell execution:} instructions invoke the script
through a shell command.
\item \textbf{Filesystem read:} the script reads the local report.
\item \textbf{Credential access:} the script retrieves an API token.
\item \textbf{Network egress:} the script sends an HTTP request
to an external service.
\end{itemize}

\textbf{Review focus.}
Check the report's sensitivity, the token's privileges, and
whether the destination is authorized. Credential access and
network egress warrant joint review; their presence alone does
not establish malicious intent.

\medskip
\textbf{Evidence and limits.}
Static evidence only; no runtime execution, threat verdict,
or exposure score is supplied for this illustration. Actual
reach depends on the session's permissions and isolation.
No attestation has been issued.
\end{minipage}}
\caption{Hypothetical presentation of a capability profile.
The example illustrates evidence, review questions, and
uncertainty; it does not demonstrate a profile-generation
or attestation system.}
\label{fig:illustrative_profile}
\end{figure}

\section{Conclusions}
\label{sec:conclusions}

Installing an agent skill grants authority, and this study measures how much. The value of the measurement lies not in any single estimate, but in the gap between two questions a marketplace can ask. The ecosystem currently asks only the narrower one. Asking whether a listing is hostile inspects the smaller population. Every
listing carrying a threat verdict also sits at medium exposure or above, so on this corpus intent implies exposure; the reverse fails badly, and at the top two exposure levels the benign listings outnumber the flagged ones four to one. That larger population is not captured by a malicious-intent verdict; its defining measured property is exposure rather than observed malice. It includes over-broad bundles carrying no observed malicious-intent verdict, which a review tuned only to intent would pass through untouched.

The ecosystem’s current permission mechanism does not support that second assessment.
Roughly one listing in eight whose frontmatter parses pre-approves any tool at all; most declarations
that exist understate the bundle behind them, and the vocabulary has no term for
reading credentials or the environment, so the two classes a reviewer would most
want named are the two that cannot be. Where a class can be named, the error follows its shape: concrete classes are over-requested in the familiar way, and
diffuse ones are missed. Uptake does not predict accuracy either, and the two
marketplaces separate in opposite directions on exactly that point.

Popularity tracks exposure upward on both marketplaces and tracks threat on neither. The two marketplaces
are ordered differently by the two axes, so a single-axis comparison ranks them
on the axis and not on their inventories. And neither analysis phase
subsumes the other: of the listings either phase flags on the risk axis, fewer
than a third are seen by both.

A marketplace can estimate a bundle’s capability set statically, at submission, before anything runs, because that is what the static phase does here and it requires no execution.
The reviewable quantity is the combination, not the count: the 397 listings holding both credential access and network egress, and the 628 satisfying the lethal trifecta, are populations small enough to read by hand and specific enough to justify the reading. Gating on those combinations at the publication boundary could prioritize review of the exposure measured here. Gating only on a declaration, an intent verdict, or an install count would miss much of it. Together, the two phases provide complementary views of a skill’s attack surface: static analysis estimates the authority available from the bundle, while runtime analysis records the subset exercised during a sandboxed session.

\section{Future Work}
\label{sec:future}

Seven of the gaps this study leaves open are specific enough to name.

\textbf{A vocabulary that can express what matters.} The format cannot declare
credential or environment access, and proposals to add exactly those fields are
open in the standard's own repository~\cite{agentskills2026proposals}. If any
of them ships, the uptake and accuracy measured here become a baseline. Whether
extending the vocabulary changes what authors declare, or only what they could
declare, is an empirical question that cannot be asked until the field exists.

\textbf{Presenting capability profiles at installation.} A further step would be to present an independently generated capability profile to users when they install a skill, including the capabilities inferred from the bundle and the evidence and uncertainty behind them. Whether such a profile helps users or marketplace reviewers make better decisions, and how its usefulness compares with repeated permission prompts, requires evaluation.

\textbf{Coverage of what a session exercises.} Detonation observes only the code
paths one run happens to take, which is why this paper reads what a listing was
seen to exercise as a lower bound. Driving each listing through several sessions, or through
inputs chosen to reach paths a single task does not, would tighten that bound and would test whether the disagreement between the phases in Section~\ref{sec:static-vs-dynamic} narrows or holds.

\textbf{Monitoring after installation.} Pre-installation profiles would also leave behavior after installation unobserved. Future work should examine whether monitoring tool use during ordinary agent sessions identifies consequential actions that a sandbox run misses, and whether such monitoring can support intervention without disrupting legitimate tasks. Repeated installation warnings are unlikely to provide a sufficient control on their own, given evidence that users habituate to security and permission warnings~\cite{vance2018tuning}.

\textbf{Code nothing invokes.} Section~\ref{sec:method-limits} reports a family of listings bundling a setup script that writes persistent agent configuration while no skill file references it. Detonation never runs such a script, so the only phase that could reach it is the one reading source rather than behavior, and no equivalent detector fires there today. Whether a static detector for
configuration-writing scripts identifies them reliably, and what a corpus-wide sweep with one would return, is untested here.

\textbf{Curation separated from catalog composition.} The marketplace
comparison in Section~\ref{sec:risk} cannot distinguish a curation effect from a difference in what each marketplace holds, because the risk indicator and the
share of a catalog that is inert prose are close relatives on this corpus. Conditioning the comparison on listings that carry any exposure at all would separate the two, and would say whether gating changes what is published or only
what is published alongside it.

\textbf{A marketplace observed over time.} Section~\ref{sec:risk} reports that the threat ordering between the two marketplaces changed when a download-banded
cohort was added to the corpus, while the risk ordering held. One snapshot cannot say which of those is the stable reading. Observing a marketplace repeatedly would measure how quickly a catalog's exposure profile moves, and whether an ordering taken from a single crawl survives the next one.

\section*{Conflict of Interest}

The scanner whose output this paper analyzes is built and operated by Metano Labs. No third party evaluated it, and no external ground truth
was available against which to validate its labels. Every rate reported here is
therefore the output of a system its authors have an interest in, and the
adjudication and reproducibility measures of
Section~\ref{sec:method-tiers} and Section~\ref{sec:method-limits} are the only
checks placed on it.

\section*{Data Availability}
\label{sec:dataavailability}
The underlying scan reports and scanned skill bundles are not publicly released. On reasonable request to the corresponding author, the authors may provide a limited, disclosure-reviewed dataset comprising selected aggregate tables underlying the paper’s figures and a de-identified per-listing exposure index. The index would contain an opaque listing identifier, marketplace, exposure level, and exposure score. Access and the scope of any release depend on a review of the requested fields for re-identification, third-party content, and security risks. Consequently, access is not guaranteed, and the complete scan results cannot be independently reconstructed from the materials offered.

Listing names, owners, repository links, and the mapping between opaque identifiers and listings will not be provided. Per-listing threat verdicts are withheld because they are thresholded model scores and adjudication of flagged listings is incomplete, as described under the Ethics Section. Raw scan reports are withheld because they may contain listing identities, third-party content, sensitive information, and security-relevant infrastructure references. The production implementation, internal orchestration, rubrics, and prompt templates are proprietary and are not available. The paper describes the measurement design and validation procedures needed to interpret its reported aggregate findings.

\section*{Ethics}
The detonation was performed in isolated sandboxes under a network policy with no open egress, and three skill-loaded runs produced attempted outbound activity classified as exfiltration. Bundles were treated as untrusted input throughout and were never executed outside that environment.

In this paper, no individual listing is named. Flagged threat verdicts are thresholded model scores rather than findings of fact, and the organization’s data-handling policy withholds per-listing threat verdicts pending adjudication. Threat outcomes are therefore reported only in aggregate, while limited de-identified exposure measurements may be shared under the conditions stated in Data Availability. Rates are reported per marketplace, and we draw no conclusion from them about either marketplace’s security program.

Two findings concern parties outside the study. Disclosure to them is the authors' responsibility and had not been completed when this paper was prepared, so no timeline is claimed here, and the findings are described generically. The
flagged listings had not been reported to either marketplace, since the data-handling policy withholds them pending adjudication, and that adjudication
is incomplete.

\bibliographystyle{IEEEtran}
\bibliography{bib/references}

\end{document}